\documentclass{ws-procs961x669}            
\begin{document}

\title{Black holes at a crossroads: late-stage evaporation in quadratic gravity}

\author{A. Bonanno}

\address{INAF, Osservatorio Astrofisico di Catania, Via S. Sofia 78, 95123 Catania, Italy\\
INFN, Sezione di Catania, Via S. Sofia 64, 95123 Catania, Italy}

\author{S. Silveravalle$^*$}

\address{SISSA - International School for Advanced Studies, Via Bonomea 265, 34136 Trieste, Italy\\
INFN, Sezione di Trieste, Via Valerio 2, 34127 Trieste, Italy\\
IFPU - Institute for Fundamental Physics of the Universe, Via Beirut 2, 34151 Trieste, Italy\\
${}^*$ E-mail: ssilvera@sissa.it}

\begin{abstract}
In General Relativity black hole evaporation leads to sudden bursts of energy and loss of information. It can be argued that these phenomena happen in the final stages of evaporation, where the semiclassical approximation needs to be refined with quantum corrections also for the gravitational interaction. A natural way to describe gravity at high energies is to add quadratic curvature terms to the Einstein-Hilbert action, i.e. quadratic gravity. At the cosmological level it is known that its classical solutions can give rise to a model of inflation that matches observations strikingly well, while in an astrophysical context it allows for the possibility of non-Schwarzschild black holes at small masses. These solutions have very peculiar properties, due to the presence of a massive spin-2 particle corresponding to a ghost at the quantum level. The branch of non-Schwarzschild solutions crosses the one of Schwarzschild ones at a specific mass which could be between the one of an asteroid and the Planck mass, depending on the value of a slightly constrained free parameter. By analyzing their dynamical stability and thermodynamical properties, we investigate what could happen to a black hole that has evaporated to this crossing point. While this investigation might not solve the problem of the endpoint of evaporation, it can shed light on the directions it might take during its last moments.
\end{abstract}


\bodymatter

\section{Introduction}\label{aba:sec1}

Our understanding of black holes as physical objects was completely renewed by the derivation of their thermodynamical properties by Gibbons and Hawking in the '70s \cite{Hawking:1975vcx, Hawking:1976de, Gibbons:1976ue}. By studying quantum fields on a classical curved spacetime, they not only understood that black holes have a mutual interaction with the surrounding environment, but they also made the first robust prediction of a semiclassical theory of gravity. Nevertheless, they quickly realized that black hole evaporation will eventually lead to a loss of information and a breakdown of the unitarity of the theory \cite{Hawking:1976ra}. This black hole information paradox was approached either by accepting the loss of information as Hawking did \cite{Hawking:1976ra}, by considering the possibility of remnants that store the missing information \cite{Chen:2014jwq}, or by rejecting the possibility of a non-unitary evaporation \textit{ab initio} \cite{Page:1979tc}.

With a conservative mindset, it can be argued that the information paradox is relevant only in the final stages of evaporation, where the temperature of radiation diverges and the semiclassical approximation needs to be refined with the inclusion of quantum corrections also for the gravitational part of the theory. Looking at the form of the one-loop counterterms of General Relativity \cite{tHooft:1974toh}, at the infrared limit of fundamental theories of gravity such as String Theory \cite{Zwiebach:1985uq}, and at non-perturbative analysis of the Renormalization Group flow of gravity \cite{Benedetti:2013jk}, it seems sensible to consider the addition of quadratic curvature terms to the Einstein-Hilbert action as first quantum corrections. The most general quadratic action can be written as
\begin{equation}\label{action}
\mathcal{S}=\int\mathrm{d}^4x\sqrt{-g}\left[\gamma R-\alpha\, C^{\mu\nu\rho\sigma}C_{\mu\nu\rho\sigma}+\beta R^2+\chi\, \mathcal{G}\right],
\end{equation}
where $\mathcal{G}$ is the Gauss-Bonnet combination, which does not contribute to the equations of motion in four dimensions, $R$ is the Ricci scalar, and $C_{\mu\nu\rho\sigma}$ is the Weyl tensor. At the quantum level this theory is renormalizable, but at the price of having a ghost particle with negative energy states which eventually spoils the unitarity of the theory \cite{Stelle:1976gc}. However, the classical solutions of this theory are good candidates to be first-order corrections of the solutions of General Relativity (note that with a FLRW ansatz for the metric the action (\ref{action}) is equivalent on-shell to the Starobinsky one \cite{Starobinsky:1980te}), and are indeed optimal as background for a refined semiclassical approximation.

In addition to various exotic solutions \cite{Lu:2015psa,Podolsky:2019gro,Silveravalle:2022lid,Bonanno:2022ibv,Daas:2022iid}, both Schwarzschild and two branches of non-Schwarzschild black holes have been found \cite{Lu:2015cqa} and characterized \cite{Bonanno:2019rsq}. Moreover, it is possible to define thermodynamical properties for non-Schwarzschild black holes, which satisfy the first law of Thermodynamics. However, as a signature of the presence of a ghost particle at the quantum level, most of the non-Schwarzschild black holes have negative mass and negative entropy. In this paper we will focus on the crossing point of Schwarzschild and non-Schwarzschild black holes, which occurs at a specific and small value of the total mass of the solutions. In particular, we will investigate the different possibilities for a black hole that has reached such a small mass in its late stages of evaporation.

\section{Black holes in quadratic gravity}

Let us consider a static, spherically symmetric ansatz for the metric in Schwarzschild coordinates
\begin{equation}\label{ansmet}
\mathrm{d}s^2=-h(r)\mathrm{d}t^2+\frac{\mathrm{d}r^2}{f(r)}+r^2\mathrm{d}\Omega^2.
\end{equation}
It can be proved that for asymptotically flat black holes the Ricci scalar is identically zero, and then the Ricci scalar squared term in the action does not contribute to the equations of motion (e.o.m.) \cite{Nelson:2010ig,Lu:2015cqa}. The e.o.m. are therefore equivalent to the ones of the Einstein-Weyl theory, which can be recast as two second-order ordinary differential equations \cite{Lu:2015cqa,Bonanno:2019rsq}. Even though the equations are much simpler, solving them analytically in their full non-linear form is impossible, and numerical methods and/or approximations have to be used. At large distances the solutions can be described by the weak field expansion \cite{Stelle:1977ry,Bonanno:2019rsq}
\begin{equation}\begin{split} \label{linsol}
h(r)=\, &1-\frac{2\,M}{r}+2S^-_2 \frac{e^{-m_2\, r}}{r}, \\
f(r)=\, &1-\frac{2\,M}{r}+S^-_2 \frac{e^{-m_2\, r}}{r}(1+m_2\, r) ,
\end{split}\end{equation}
where $m_2^2=\gamma/2\alpha$ is the mass of the ghost particle of the quantum theory, $M$ is the total ADM mass of the solution (in units of $G$), and $S_2^-$ is a free parameter that we will call ``Yukawa charge'' from now on. We note that the Yuakawa term in the Newtonian potential $\phi(r)=\frac{1}{2}\left(h(r)-1\right)$ will be either attractive or repulsive if the Yukawa charge is respectively either negative or positive. The metric close to the horizon radius $r_H$ can instead be written as a series expansion
\begin{equation}\begin{split} \label{expsol}
h(r)=\, &h_1(r-r_H)+h_2(r-r_H)^2+..., \\
f(r)=\, &f_1(r-r_H)+f_2(r-r_H)^2+...,
\end{split}\end{equation}
with only $h_1$, $f_1$ and $r_H$ as free parameters \cite{Lu:2015cqa}. The thermodynamical properties of the black hole can be expressed in terms of these free parameters as
\begin{equation}\begin{split} \label{therm}
T_{BH}=\, &\frac{\sqrt{h_1 f_1}}{4\pi}, \\
S_{BH}=\, &16\pi^2\gamma\left(r_H^2+\frac{2}{m_2^2}\left(1-f_1 r_H\right)\right),
\end{split}\end{equation}
where the entropy has been defined using the Wald definition \cite{Wald:1993nt,Fan:2014ala}, and with $k_B=1$.

To study black holes we use the shooting method to numerically integrate the equations of motion between the two boundaries (\ref{linsol}) and (\ref{expsol}); with such procedure, we can simultaneously find the values of the parameters $(M,S_2^-,h_1,f_1)$ for any fixed radius $r_H$. A detailed procedure description can be found in previous literature \cite{Bonanno:2019rsq,Silveravalle:2023lnl}. During the calculations all the quantities have been adimensionalized using the mass $m_2$ as energy unit, and its inverse as length unit. Moreover, the entropy in (\ref{therm}) has been divided by $16\pi\gamma$ (\textit{i.e.} multiplied by $G$) to be consistent with the mass parameter $M$ in (\ref{linsol}). 

\begin{figure}[hbt]
\begin{center}
\includegraphics[width=\textwidth]{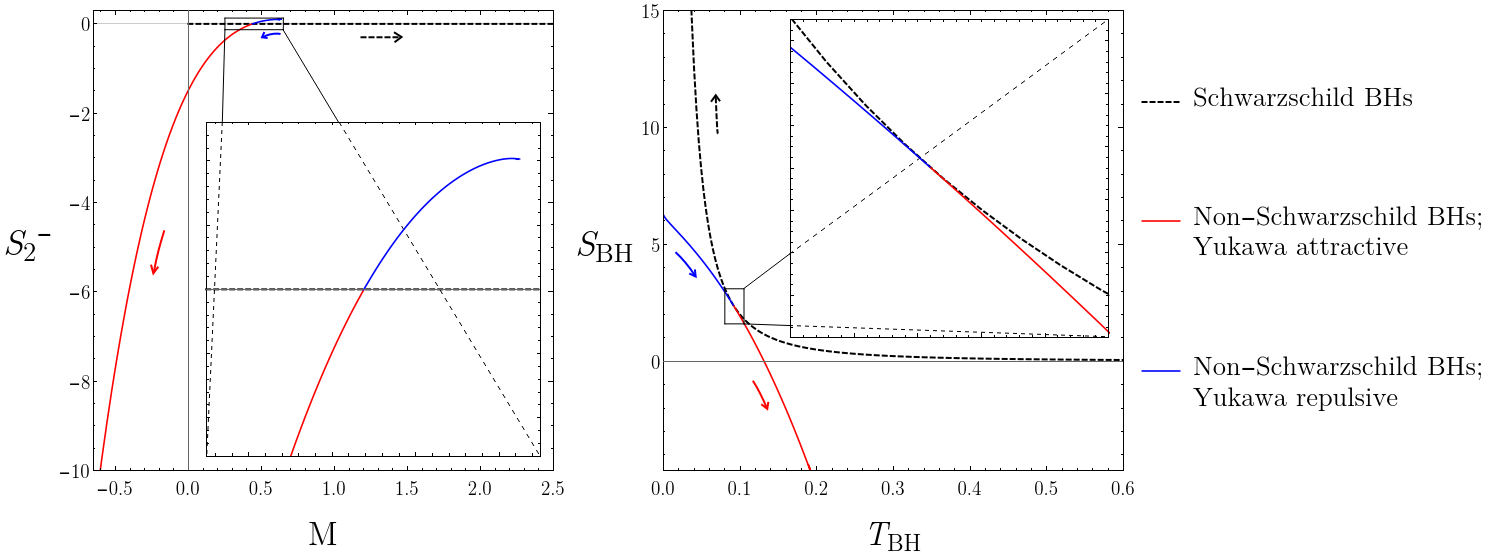}
\end{center}
\caption{Properties of black holes in quadratic gravity with the gravitational parameters at large distances on the left panel and the thermodynamical properties on the right panel; the arrows indicate the directions of increasing horizon radius.}
\label{fig1}
\end{figure}

As stated in the literature\cite{Lu:2015cqa,Goldstein:2017rxn,Bonanno:2019rsq}, and shown in Fig. \ref{fig1}, together with the standard Schwarzschild solution there are two branches of non-Schwarzschild black holes: Yukawa repulsive black holes (in blue in Fig. \ref{fig1}) which are present only for a small range of values of the parameters $(M,S_2^-)$ and of Wald entropy, and they have small horizon radius and temperature; and there are Yukawa attractive black holes (in red in Fig. \ref{fig1}) which instead have unbounded values of the parameters $(M,S_2^-)$ and of Wald entropy, and they have arbitrarily large horizon radius and temperature. We note that most of Yukawa attractive black holes have negative mass and entropy, which can be interpreted as the classical manifestation of the presence of a ghost particle at the quantum level. 

The crossing point of the two families of black holes is at a horizon radius of $r_H\simeq 0.876$ in our units. Restoring physical units we have the transition happening at the values shown in Tab. \ref{tab1}; in particular in the fourth column we considered the maximum value $\alpha<10^{60}$ allowed by torsion tests\cite{Giacchini:2016nta}, and in the fifth a realistic value $\alpha\sim 10^8$ which is of the order of the free parameter of the Starobinsky model of inflation. Also for the extremal values of the parameters the corresponding black holes are too small to be astrophysical objects, but it is even more clear with the realistic values, where the solutions are almost of Planckian size. It is then sensible to expect astrophysical processes to generate Schwarzschild black holes, and consider the crossing point only relevant in the late stages of black hole evaporation.

\begin{table}[htb]
\tbl{Adimensional, dimensionful, the maximum and a realistic value for the main physical properties of the black hole at the crossing point.}
{\begin{tabular}{@{}ccccc@{}}
\toprule
Characteristics  & Adimensional value & Dimensionful value & Maximum value & Realistic value\\\colrule
$r_{H,tr}$ & 0.876 & $1.4\sqrt{\alpha}10^{-37}km$ & $1.4\cdot 10^{-7}km$ & $1.4\cdot 10^{-33}km$\\
$M_{tr}$ & 0.438 & $4.8\sqrt{\alpha}10^{-38}M_\odot$ & $4.8\cdot 10^{-8}M_\odot$ & $4.8\cdot 10^{-34}M_\odot$ \\
$T_{BH,tr}$ & 0.091 & $1.3\alpha^{-1/2}10^{30}K$ & $1.3 K$ & $1.3\cdot 10^{26}K$ \\\botrule
\end{tabular}}
\label{tab1}
\end{table}

\subsection{Classical perturbations and stability}\label{sub:per}

A fundamental tool to study black hole solutions is the behavior of classical linear perturbations, which is closely related to their stability. It has been shown\cite{Held:2022abx} that classical perturbations can be decomposed into three modes, that is a massless spin-2, a massive spin-0 and a massive spin-2, precisely as in the quantum case\cite{Stelle:1976gc}. The only mode with a non-trivial behavior is the massive spin-2 one, which corresponds to the ghost particle at the quantum level. In our case the degrees of freedom of the perturbation can be reduced to the dynamics of a single scalar function satisfying a Regge-Wheeler-Zerilli-like equation
\begin{equation}\label{zeril}
\left(\frac{\mathrm{d}^2}{\mathrm{d}t^2}-\frac{\mathrm{d}^2}{\mathrm{d}{r^*}^2}\right)\varphi(r,t)+V(r)\varphi(r,t)=0,
\end{equation}
where $r^*$ is the tortoise coordinate and $V(r)$ is a potential that depends on the background metric (the full form of the potential can be found on the original paper of Held and Zhang\cite{Held:2022abx}).

\begin{figure}[hbt]
\begin{center}
\includegraphics[width=\textwidth]{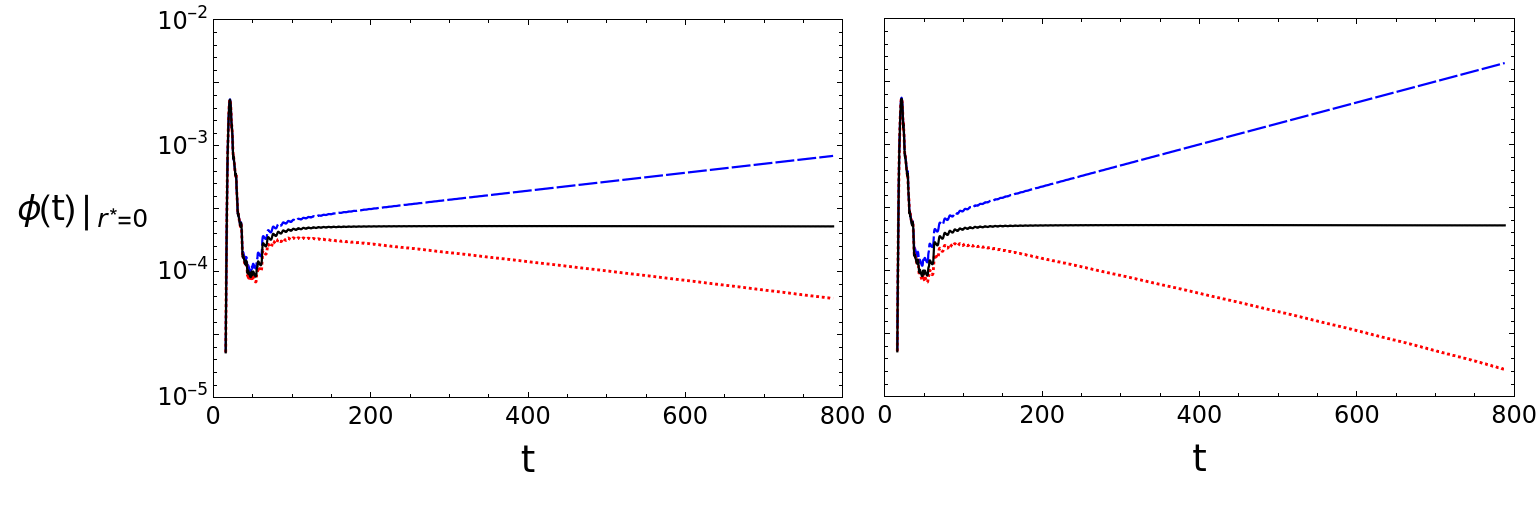}
\end{center}
\caption{Moving average over an interval $t_{MA}=25$ of the time evolution of perturbations at $r^*=0$, for black holes with horizon radius $r_H=0.874$ in dashed blue, $r_H=0.876$ in solid black and $r_H=0.878$ in dotted red; in the left panel there are Schwarzschild black holes and in the right one there are non-Schwarzschild black holes.}
\label{fig2}
\end{figure}

We numerically integrated equation (\ref{zeril}) on a light-cone coordinates grid, and found that black holes with an event horizon larger than the one at the crossing point, that is large Schwarzschild black holes and Yukawa attractive black holes, have exponentially suppressed perturbations and are linearly stable, while smaller black holes, that is small Schwarzschild black holes and Yukawa repulsive black holes, have exponentially increasing perturbations and are linearly unstable; at the crossing point the massive spin-2 perturbation has zero imaginary frequency, and therefore lead to unsuppressed oscillations. This integration confirms previous ones on the calculations of specific frequencies\cite{Lu:2017kzi,Held:2022abx}, without assuming any mode decomposition in the time domain. In Fig. \ref{fig2} we show the time evolution of these perturbations in the vicinity of the crossing point; to have a clear picture we show the perturbation at $r^*=0$, in order to avoid the exponential suppression in the radial direction due to the massive nature of the perturbation, and we show the moving average to remove the oscillations leaving only the information on their exponential behavior. In particular, we note that the Yukawa repulsive black hole has an exponential growth much larger than the Schwarzschild black hole with the same horizon radius.

\section{Black hole crossroads}

As argued before, we expect astrophysical or cosmological processes to generate Schwarzschild black holes. Thanks to the sanity check of classical perturbations we know that large Schwarzschild black holes are stable against all types of perturbations, and therefore their evolution will be determined by the evaporation. In Hawking's picture the time evolution of evaporation can be inferred by the flux equation 
\begin{equation}\label{flux}
\mathcal{H}_{tr}=\frac{1}{2}\langle T_{tr}\rangle,
\end{equation}
where $\langle T_{\mu\nu}\rangle$ is the expectation value of the stress-energy tensor of radiation and $\mathcal{H}_{\mu\nu}$ is the e.o.m. tensor. To be consistent, it is necessary to adopt an adiabatic approximation in which the metric keeps its original form, but with time-dependent parameters \cite{Hawking:1975vcx,Gregory:2021ozs}. Within this approximation equation (\ref{flux}) can be written explicitly for a Schwarzschild black hole as
\begin{equation}\label{evapeq1}
\frac{\mathrm{d}M}{\mathrm{d}t}=-\displaystyle\sum_{s,l,m}\int\frac{\mathrm{d}\omega}{2\pi}|{}_s\Gamma_{\omega lm}|^2\frac{\omega}{\mathrm{e}^{\omega/T_{BH}}-1},
\end{equation}
where $|{}_s\Gamma_{\omega lm}|^2$ is the grey-body factor for a perturbation of spin $s$, frequency $\omega$, and $l$ and $m$ are the degree and order of spherical harmonics. If, moreover, we consider a high-frequency limit, for which $|{}_s\Gamma_{\omega lm}|^2\to C_s$, the equation can be greatly simplified to  
\begin{equation}\label{evapeq2}
\frac{\mathrm{d}M}{\mathrm{d}t}\sim -\tilde{\sigma}T_{BH}^2,
\end{equation}
where $\tilde{\sigma}$ is a number that takes into account the relative weights of perturbations of different spin $C_s$. Using the mass-temperature relation of the Schwarzschild metric $T_{BH}=1/8\pi M$, equation (\ref{flux}) can then be solved analytically as
\begin{equation}\label{schwaevo}
M(t)=M_0\left(1-\frac{t}{t_e}\right)^{1/3},
\end{equation}
where $M_0$ is the initial mass, and the relative weights of perturbations of different spin are encoded in the evaporation time $t_e$. At time $t_e$ the black hole is indeed completely evaporated, and the information paradox occurs. However, we consider the black hole to evaporate following equation (\ref{schwaevo}) only until a time $t_{tr}$ where it has reached the mass $M_{tr}$, that is the black hole crossroads.

\subsection{First direction: standard evolution}

At first, let us consider the possibility that there is no transition between Schwarzschild and non-Schwarzschild black holes, and a large Schwarzschild black hole evaporates in a small Schwarzschild black hole. On a Schwarzschild background the equation for a tensor perturbation $\psi_{\mu\nu}$ take the rather simple form\cite{Lu:2017kzi}
\begin{equation}\label{pertschwaeq}
    \Box \psi_{\mu\nu}+2R_{\mu\rho\nu\sigma}\psi^{\rho\sigma}-m_2^2\psi_{\mu\nu}=0.
\end{equation}
At large distances equation (\ref{pertschwaeq}) can be solved analytically for a mode with frequency $\omega$; considering as an example the $tt$ component of the metric we find that
\begin{equation}\label{pertschwa}
g_{tt}\sim \bar{g}_{tt}+\psi_{tt}\sim 1-\frac{2\,M}{r}+C\,\mathrm{e}^{-i\omega t}\frac{\mathrm{e}^{-\sqrt{m_2^2-\omega^2}\,r}}{r}.
\end{equation}
However, as shown in subsection \ref{sub:per}, the perturbations around small Schwarzschild black holes have positive imaginary frequency and grow exponentially. This means that even an infinitesimal perturbation will generate a relevant Yukawa term in the gravitational potential. In particular, we note that at the transition point it is possible to have static, that means unsuppressed, perturbations with zero frequency; the perturbed metric then takes the form
\begin{equation}\label{pertschwa2}
g_{tt}\sim \bar{g}_{tt}+\psi_{tt}\sim 1-\frac{2\,M}{r}+C\,\frac{\mathrm{e}^{-m_2\,r}}{r},
\end{equation}
which is precisely the one of non-Schwarzschild black holes at large distances. If we try to follow the small Schwarzschild direction, we are then pushed back to non-Schwarzschild solutions by perturbations. Recently it has been found some evidence of this behavior also at the non-linear level in a paper by East and Siemonsen\cite{East:2023nsk}, where it has been found that an unstable small Schwarzschild black hole will either stabilize as a black hole with a larger horizon radius or vanish in a finite amount of time, depending on the sign of the initial condition for the perturbation. In particular, we note that while in their paper East and Siemonsen talk about increasing and vanishing mass, they used Christodoulou's formula for the mass which is proportional to the horizon radius of the black hole. We then interpret their result as a small Schwarzschild black hole either stabilizing as a large stable Yukawa attractive one or following the unstable route of Yukawa repulsive ones, depending on the sign of the $C$ parameters on equations (\ref{pertschwa}, \ref{pertschwa2}). The first direction of an evolution that avoids non-Schwarzschild black holes seems excluded by the dynamics of perturbations.

\subsection{Second direction: standard evaporation}

A purely thermodynamical description of evaporation requires the system to be in dynamical equilibrium. This means that an evaporating Schwarzschild black hole that has reached the transition point should follow the branch of Yukawa attractive black holes. In this context it is possible to apply the same approximations used to derive the equation describing the evaporation process in General Relativity, that is the adiabatic approximation and the high-frequency limit to the flux equation (\ref{flux}). However, to evaluate equation (\ref{flux}) with the e.o.m. of quadratic gravity, it is necessary to stress the adiabatic approximation a bit more and consider only terms linear in the first derivatives in time, and no higher-order derivatives in times. At large distances, we can use the linearized expansion (\ref{linsol}) and obtain once again equations (\ref{evapeq1}, \ref{evapeq2}). The thermodynamical properties of the black hole are guaranteed to persist throughout the evaporation since, in this stressed adiabatic approximation, it is possible to evaluate equation (\ref{flux}) also close to the horizon and obtain
\begin{equation}
T_{BH}\frac{\mathrm{d}S_{BH}}{\mathrm{d}t}\sim-\tilde{\sigma}T_{BH}^2\sim\frac{\mathrm{d}M}{\mathrm{d}t}.
\end{equation}

\begin{figure}[hbt]
\begin{center}
\includegraphics[width=\textwidth]{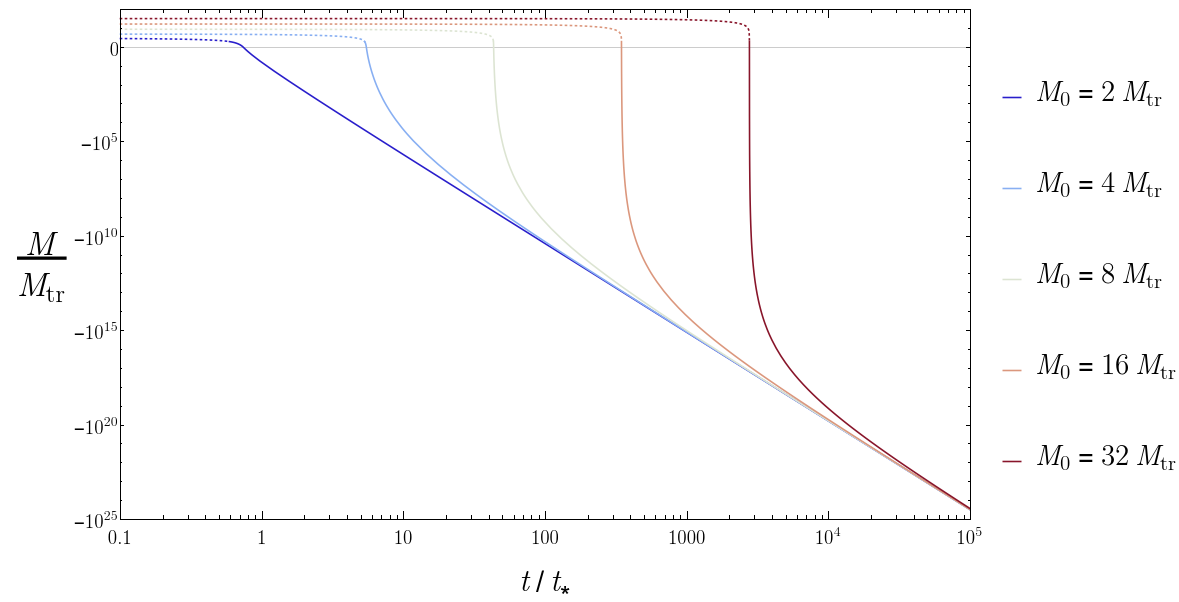}
\end{center}
\caption{Evaporation of Schwarzschild black holes of different initial mass $M_0$ which undergo a transition into the stable, Yukawa attractive, branch of non-Schwarzschild black holes; the lines are dashed before the transition point, and solid after the transition.}
\label{fig: stable evap}
\end{figure}

For a qualitative description of the time evolution of evaporation, we model the dependence of the temperature from the mass for non-Schwarzschild black holes as
\begin{equation}
    T_{BH}\sim A^{-1/2}(M_{mtp}-M)^{\eta/2},
\end{equation}
where $M_{mtp}$ is the mass of the non-Schwarzschild black hole with a vanishing horizon radius, and $A$ and $\eta$ have been obtained fitting the numerical $T(M)$ dependence, and whose best-fit values are $A=32.451$ and $\eta=0.787$. The time evolution is then
\begin{equation}\label{evol stab}
\begin{split}
M(t)\sim M_{mtp}-\left(\left(M_{mtp}-M(t_0)\right)^{1-\eta}+\frac{\tilde{\sigma}(1-\eta)}{A}(t-t_0)\right)^{\frac{1}{1-\eta}},
\end{split}
\end{equation}
with mass decreasing as the evaporation proceeds. However, a crucial difference with the evaporation of Schwarzschild black holes is that \emph{there is no minimum mass}, and black holes can evaporate to negative masses arbitrarily large in modulus. If we consider a Schwarzschild black hole with initial mass $M_0$ which reaches the transition point at time
\begin{equation}
    t_{tr}=\frac{64\pi^2}{3\tilde{\sigma}}\left(M_0^3-M_{tr}^3\right),
\end{equation}
we can estimate its present mass. Figure \ref{fig: stable evap} shows the evaporation process of Schwarzschild black holes with different initial masses; in particular we show the total mass in units of the transition point mass $M_{tr}$, in function of the reduced time $t_*=10^5\alpha^{3/2}t_P$, where $t_P$ is Planck time. It is manifest that after some time all the black holes share a common evaporation behavior. 

While it is already clear that this evaporation could lead to catastrophic consequences, it is instructive to restore units with a more intuitive physical meaning. Let us consider a primordial Schwarzschild black hole of mass $M<10^{11}Kg$ generated in the early universe $t\sim 10\, Gy$: in the standard picture such a small black hole would have been completely evaporated at present time, while in this model, for values of $\alpha<10^{37}$, it would have reached the transition point and started evaporating as a Yukawa attractive black hole. However, for such values of $\alpha$ the present mass would be $M<-10^{16}M_\odot$, and for values of $\alpha$ of the order predicted by CMB anisotropies for Starobinski’s model, that is $\alpha\sim 10^8$, it would be $M<-10^{206}M_\odot$, that is $-10^{108}$ times the estimated total energy of the universe. It is then clear that for an anthropic principle the transition of an evaporating Schwarzschild black hole into the stable Yukawa attractive branch has to be excluded, as it is incompatible with our universe. 

\subsection{Third direction: ghost instability}

The only direction left for evaporation seems to be the one of unstable Yukawa repulsive black holes. However, we believe this is not only the last possibility left, but the most sensible direction for the evolution of black holes. Let us consider a black hole at the transition point: the Hawking process will create all possible types of particles, including the massive spin-2 quantum ghosts. These particles will satisfy the equations of motion of classical perturbations, and are therefore subject to the same instabilities; these instabilities will force a change in the expectation value for the vacuum and then a change in the background metric. As shown in Fig. \ref{fig2}, Yukawa repulsive black holes have perturbations that grow faster than the corresponding Schwarzschild black holes: we believe that this is an indication that the quantum random fluctuations at the transition point will trigger a classical transition into this branch of black holes. Moreover, if quantum ghosts truly start dominating the evaporation process, we expect the total mass of the solution to increase rather than decrease, due to the emission of negative energy particles; this is precisely the case of Yukawa repulsive black holes, which have a mass larger than the one at the transition point. 

Nonetheless, it is unclear what can happen to a black hole that transitions into a Yukawa repulsive black hole, having no possibility to use the adiabatic approximations anymore and being forced to use a time-dependent metric. If we assume that the instability does not change the nature of the metric and an event horizon is still present, and that the evaporation process has to end on a static solution, the natural endpoint of evaporation is the Yukawa repulsive black hole with a vanishing horizon radius. This solution is actually the massive triple point discussed in a previous paper\cite{Silveravalle:2022wij}, which is the only non-trivial point of the parameter space where all the types of solutions of quadratic gravity coincide. This solution is a naked singularity, but of exotic nature and with the possibility of being present in our universe\cite{Holdom:2016nek,Silveravalle:2023lnl}. However, a detailed discussion of the evolution through the direction of ghost instability is beyond the scope of this paper and we postpone it to future work. 

\section{Conclusions}

In this paper we investigated the different possible directions for the evaporation of a black hole in quadratic gravity. In this theory there is a crossing point at very small masses between Schwarzschild black holes and two branches of non-Schwarzschild ones, which is reached by astrophysical Schwarzschild black holes in the late stages of evaporation. The possible directions at this crossing point are: evaporating into small Schwarzschild black holes, evaporating into large Yukawa attractive black holes, and evaporating into the small and unstable Yukawa repulsive black holes. The first direction is excluded by the dynamics of perturbations: it is manifest at the linear level, and seemingly confirmed at the non-linear level, that massive spin-2 perturbations will force a transition of Schwarzschild black holes into non-Schwarzschild ones. The second direction is excluded by an anthropic principle: Yukawa attractive black holes have negative specif heat as the Schwarzschild ones, but can have indefinitely large negative masses; evaporation through this direction would lead to extremely large and gravitational repulsive isolated solutions which are incompatible with our universe. The third direction of a transition into Yukawa repulsive black holes driven by the instability of the massive spin-2 ghost particle seems to be the only one left; nonetheless, we argue that exponentially growing modes should trigger a transition into this type of black hole. While in this paper we did not investigate deeply the evolution through this ghost instability, we argue that the endpoint might be an exotic naked singularity with physically acceptable features. In the future we aim to address this point more carefully, hoping to point out the direction for black hole evaporation at high energies.

\bibliographystyle{ws-procs961x669}

\end{document}